\documentclass[letterpaper, 12 pt, conference]{ieeeconf}  

\IEEEoverridecommandlockouts                              
\usepackage{graphicx} 
\usepackage{bm}

\title{\LARGE \bf
Blockchain Meets LLMs: A Living Survey on Bidirectional Integration}

\author{Jianghao Gong$^{}$,  Peiqi Yan$^{}$,  Yue Zhang$^{}$, Hongli An$^{}$, Logan Liu$^{}$
\thanks{}
\thanks{$^{1}$Undergraduate Research Training Program, IEEE Hainan University Student Branch}%
\thanks{$^{2}$Blockchian Group, RobAI-Lab@HainanU}%
}

\begin{document}

\maketitle
\thispagestyle{empty}
\pagestyle{empty}

\begin{abstract}
In the domain of large language models, considerable advancements have been attained in multimodal large language models and explainability research, propelled by the continuous technological progress and innovation.   Nonetheless, security and privacy concerns continue to pose as prominent challenges in this field.   The emergence of blockchain technology, marked by its decentralized nature, tamper-proof attributes, distributed storage functionality, and traceability, has provided novel approaches for resolving these issues.   Both of these technologies independently hold vast potential for development;   yet, their combination uncovers substantial cross-disciplinary opportunities and growth prospects.   The current research tendencies are increasingly concentrating on the integration of blockchain with large language models, with the aim of compensating for their respective limitations through this fusion and promoting further technological evolution.   In this study, we evaluate the advantages and developmental constraints of the two technologies, and explore the possibility and development potential of their combination.   This paper primarily investigates the technical convergence in two directions: Firstly, the application of large language models to blockchain, where we identify six major development directions and explore solutions to the shortcomings of blockchain technology and their application scenarios;   Secondly, the application of blockchain technology to large language models, leveraging the characteristics of blockchain to remedy the deficiencies of large language models and exploring its application potential in multiple fields.

\end{abstract}
\emph{Keywords: Blockchain, Large Language Models}
\section{INTRODUCTION}
 In the wave of rapid development of artificial intelligence technology, LLMs multimodal large language models are becoming a key force driving innovation in the field, thanks to their strong language understanding and generation capabilities, as well as their increasing explainability. For instance, the study “Utility of Multimodal Large Language Models (GPT-4 Vision and Large Language and Vision Assistant) in Identifying Melanoma Across Different Skin Tones” \cite{cirone2024assessing} demonstrates the application of these models in melanoma detection.Similarly, language models such as GPT-4 have been used to explain the functions of neurons in similar models \cite{bills2023language}. Meanwhile, models such as GIT-Mol \cite{liu2024git} and Gemini \cite{team2023gemini} have successfully integrated graph, image, and text information, showcasing their outstanding capabilities in cross-modal processing. These models have demonstrated exceptional performance in tasks such as image, audio, video, and text understanding. 
  
  Despite their vast potential for applications, LLMs also face numerous challenges, particularly in data protection and privacy security. One particularly noteworthy issue is that LLMs are highly susceptible to manipulation, and may inadvertently reveal personal identity information (PII) in the process. For example, there have been cases where LLMs, under improper guidance, disclosed sensitive personal information, including full name, residence address, email, phone number, and related Twitter accounts to unauthorized users \cite{carlini2023extracting}. While LLMs can generate seemingly precise responses, they actually do not understand the language they are using, and may have absorbed incorrect information and biases during training, making it difficult to determine the authenticity of the content. As probabilistic algorithms, the reliability of their output is also questionable, and therefore, human oversight is needed to prevent the spread of incorrect information. Due to the lack of proper regulation and responsible use, the spread of incorrect information has become a problem \cite{harrer2023attention}. Meanwhile, the widespread phenomenon of hallucination in LLMs highlights the disparity between the generated content and verifiable real-world facts, often manifesting as inconsistencies or fabrications \cite{huang2023survey}. These problems have not yet been fully resolved, and effective mitigation measures are relatively limited.
  
  It is evident that to address the problems arising from large language models and prevent the emergence of more vulnerabilities, new technologies need to be introduced into the field. As a rapidly developing technology, blockchain's decentralized structure, distributed ledger, consensus algorithm, smart contracts, and asymmetric encryption ensure the security and transparency of the network, making it an innovative and reliable digital shared ledger that can record and store immutable data and transactions. It has shown disruptive potential in multiple industries such as finance, healthcare, and supply chain management \cite{dutta2020blockchain}. The characteristics of blockchain systems make it an ideal platform for enhancing the robustness of large language models. This integration provides the necessary framework, making it possible to incorporate more advanced privacy protection measures, enhanced logical reasoning verification, adversarial attack defense techniques, and similar safeguards into the design of large language models.
  
  The blockchain is not yet a fully developed technology, and it still faces many problems, the most notable of which are computational attacks, network attacks, and smart contract security issues. Computational attacks, such as the 51\% attack, allow attackers to manipulate transaction records by controlling most of the computing power, endangering the immutability of the blockchain. Network attacks, such as DoS attacks, impede normal communication between nodes and affect the stability of the blockchain network. Security vulnerabilities in smart contracts may result in financial losses and impaired contract functionality \cite{hazra2022blockchain}
. Meanwhile, blockchain also faces significant scalability challenges, as data storage becomes an increasingly important challenge as the blockchain network grows, especially in memory-constrained applications such as the Internet of Things. The original trilemma of scalability, decentralization, security, and trust has expanded into a four-fold challenge, involving scalability, decentralization, security, and trust. Currently, blockchain systems are difficult to satisfy all four aspects simultaneously. For example, private and consortium chains may make progress in enhancing security and scalability, but this is often at the expense of decentralization; while blockchain based on DAG (directed acyclic graph) may perform well in enhancing scalability and decentralization, but may have insufficient security and trust-building capabilities \cite{sanka2021systematic}.

However, this challenge also provides opportunities for the integration of blockchain technology with large language models. By combining the immutability and distributed ledger characteristics of blockchain with the intelligent processing capabilities of large language models, we can open up new solutions to alleviate or even solve these challenges. In recent years' technological development trends, the integration of blockchain technology with large language model fields has become a focus of research. As an emerging field, blockchain services for large language models is still in its early stages and requires in-depth analysis and research. Our goal is to help researchers have a more comprehensive understanding of the development trends of these two fields, especially to explore how to efficiently integrate blockchain technology into large language model systems and how large language models can help improve blockchain technology. Through this review, this review aims to provide insights to researchers to overcome the challenges that may arise in the process of technological integration and promote further development and application of technological integration.
 
\section{THE COMBINATIONS}
\paragraph{APPLY BLOCKCHAIN TO LLMs}
\subsection{ Data Management and Data Privacy Protection}
The current large language model is based on Transformer as the core framework, using a self-attention mechanism and a multi-layer structure to train strong language understanding and generation capabilities through large-scale data. A very large amount of data is required in the training iteration. However, due to personal privacy and security considerations, many data cannot be fully utilized, which seriously restricts the development of large language models. Due to the decentralization and anonymity of blockchain, the following problems can be solved by introducing it into large language models.
\subsubsection{Privacy-preserving data sharing and training using blockchain}
 The common federated learning architecture is client-server. The server transmits intermediate parameter models such as gradients, model parameters, and embedded representations during the training process to the client locally. After the client updates the model locally, it sends the parameter model back to the central server. The central server After receiving the parameters from all parties, the calculation is aggregated, and the updated model parameters are issued again. This process loops until the model converges or the training is terminated. Another common architecture is the end-to-end network architecture. This architecture consists of terminals holding data, with no central server. A terminal holding training data then performs data training and passes the model parameters to the next terminal until the model converges or the training is terminated. In this process, the blockchain can manage the communication between devices and the update of model parameters, so that the data is always saved locally, ensuring data security to a certain extent. 

\subsubsection{Safe and reliable personal data protection and sharing}
The training data of large language models may contain sensitive personal training information or private data, and unprocessed data may lead to privacy leaks. Academic research shows the risk that models may inadvertently generate text containing personal information is real. In addition, personal data is mainly stored in the cloud. However, there are certain risks in using third-party cloud storage services. Additionally, users lack ownership of their data as they may belong to third-party service providers once registered, and users have no right to know and control private data. Through the PDPChain proposed by Liang \cite{56} and others, based on a multi-center and partially decentralized consortium chain, only authorized nodes that pass the identity verification mechanism can join the blockchain network. A third-party trusted module CA is also introduced for node identity verification, certificate issuance and management. The RAFT consensus protocol is used to ensure system stability when some nodes crash. Through DAPP and smart contracts, users initiate transactions, data is encrypted and uploaded to the blockchain, and access control is managed by the CP-ABE method to ensure that only data can be accessed if conditions are met.

\subsubsection{Smart contracts and access control }
The automatic execution capability of smart contracts can store personal transaction records between users and service providers on the blockchain without the need for a trusted center, serving as a reliable legal certificate for the existence of transactions to prevent theft of personal information. criminal behavior. The actual content of personal information will not be stored on the blockchain. Only transaction records are encrypted and stored on the blockchain. Only the parties involved in the transaction can decrypt and view it, ensuring user privacy. In addition, data encryption, Technologies such as homomorphic encryption to protect data stored on the blockchain. Homomorphic encryption allows computations to be performed on encrypted data without the need for decryption, thus protecting the privacy of the data

\subsection{Decentralized model training }
Different devices in our daily lives generate and collect a lot of data, but it is difficult for a central server to collect and process all this data. With the diversification of device functions, the emergence and development of 5G and the Internet of Things (IoT) have made the generation and circulation of data more rapid and extensive, which provides more development space for distributed large language model training. Among them, federated learning is developing rapidly. However, traditional federated learning frameworks face a huge risk of  Single Point Of Failure (SPOF) due to their heavy reliance on central servers. This means that if the central server fails or is attacked, the entire learning process may be interrupted, affecting the stability and reliability of model training. In addition, it lacks incentives, resulting in insufficient contribution of local devices to global model training. Due to the lack of recognition and rewards for the contributions of participating devices, device owners may lack motivation to participate in federated learning projects, which limits the application potential of federated learning in a wider range of scenarios. As a result, researchers are exploring how to design more robust federated learning architectures and develop effective incentives to encourage wider device participation and improve overall learning efficiency.

\subsubsection{Decentralized federated learning system }
Decentralized Federated Learning  systems address the single point of failure problem in traditional federated learning through peer-to-peer communication, but they face challenges such as malicious clients, low-quality models, and the lack of incentive mechanisms. The Chain FL system utilizes the Ethereum blockchain and the authority proof consensus mechanism to improve response speed, and implements large-scale federated learning through smart contracts. 
\cite{57} Kim proposed a method that stores the global model on the blockchain and ensures the secure sharing of model updates. Ramanan et al. further used a private Ethereum network to aggregate device updates \cite{58} .The Fedstellar platform supports various federated learning configurations \cite{59}, and VDFChain provides a secure and verifiable decentralized federated learning solution based on a committee-based blockchain \cite{60}. In addition, some research has proposed a completely decentralized federated learning method, which shares knowledge through the teacher-student role allocation of the model without transmitting original data or parameters. Another study proposed a general DFL framework that balances communication and computational costs through multi-round local updates and inter-node communications, and introduces compressed communication to improve efficiency. The ProxyFL scheme improves communication efficiency and privacy protection by maintaining private and proxy models \cite{61}. These studies demonstrate the potential of DFL in enhancing the robustness and privacy of machine learning model training.

\subsubsection{Incentive mechanism design in decentralized training}
Incentive mechanisms are crucial in federated learning, as they compensate participating clients for the costs of their computational and data contributions, ensuring data diversity and the generalization ability of the model. They also promote the fairness, efficiency, and sustainability of the system, encouraging clients to actively participate in model training, prevent malicious behavior, and protect user data privacy. The incentive mechanism design problem in blockchain-based federated learning (BCFL) particularly focuses on how to effectively allocate resources between training and mining tasks \cite{62}. An innovative two-stage Stackelberg game method is proposed to optimize resource allocation strategies and ensure that clients can make reasonable decisions on computational capability allocation in the case of incomplete information \cite{63}. In addition, a new federated learning incentive mechanism, FDFL, is introduced. This mechanism provides a more comprehensive contribution evaluation and reward allocation scheme by considering the differences in data category distribution and the honesty of the server. Through security analysis and experimental verification, FDFL demonstrates excellent performance in terms of model accuracy and robustness while ensuring the truthfulness of the incentive mechanism and individual rationality \cite{64}. These studies provide new perspectives and solutions for resource management and incentive strategies in decentralized learning environments.

\subsection{Decentralized AI}

With the increasing complexity of artificial intelligence models and the rising demand for scalable, secure, and democratized AI services, the integration of artificial intelligence with distributed systems is becoming increasingly important \cite{65}. Distributed Artificial Intelligence (DAI) has become a key component of future intelligent systems due to its excellent scalability, security, and real-time response capabilities. DAI effectively handles large datasets and complex tasks through decentralized decision-making mechanisms, reduces biases, and optimizes resource allocation. The combination with blockchain technology ensures data integrity and anti-tampering capabilities, providing security for AI applications in sensitive fields such as healthcare and finance. In addition, the distributed architecture of DAI enhances the fault-tolerance of the system, and the failure of a single node does not affect the stability of the entire system. The dynamic resource allocation capability allows DAI to adapt to fluctuations in workloads, improving the utilization efficiency of computing resources. The latest research shows that DAI can expand AI systems while maintaining performance, particularly in multi-agent systems, achieving significant progress compared to traditional methods \cite{66}. 

The current technological breakthroughs in the field of Distributed Artificial Intelligence (DAI) cover key innovations such as decentralized AI markets, peer-to-peer large language models (LLMs), and optimistic machine learning on the blockchain. GradientCoin proposes a decentralized LLM framework \cite{67}, promoting user contribution of computing resources through the gradient coin incentive mechanism and proving the convergence of the training mechanism within the distributed learning system. opML enables blockchain systems to perform AI model inference through an interactive anti-fraud protocol \cite{68}, introducing fraud proof protocols and multi-stage fraud proof, enhancing economic and execution efficiency. In addition, there is a paper discusses a decentralized strategy optimization framework based on models, solving the scalable decision-making problem in large-scale network control, and SmartVM, an intelligent contract virtual machine designed for on-chain DNN computation \cite{69}, significantly improving the performance of blockchain-based AI systems. These technical contributions not only address the challenges of centralization, data privacy, and on-chain AI computation but also provide new ways to achieve safer and more efficient AI services.

\paragraph{APPLY LLMs TO BLOCKCHAIN}
\subsection{Application in smart contract auditing}
Smart contract auditing is crucial, as smart contracts cannot be changed once deployed. Auditing can discover and fix potential vulnerabilities, thereby maximizing protection of funds and preventing hacker attacks. At the same time, it helps contracts comply with laws and regulations, reducing compliance risks. Moreover, auditing can optimize performance, enhance contract efficiency, and protect the reputation of project parties, preventing reputational and financial losses due to security issues. In short, smart contract auditing is a key step in maintaining the security, compliance, and trust of the blockchain ecosystem. With the widespread application of large language models, there has been an increase in recent research on using large language models for smart contract auditing.

At present, there are directions such as the use of large language models (LLMs) for vulnerability detection, vulnerability avoidance during automatic code completion, identification of machine unauditable vulnerabilities (MUBs), development of audit frameworks, context-driven audit techniques, dataset construction, and logical reasoning and proof writing in the audit process. These methods can improve the efficiency and accuracy of smart contract audits to a certain extent. Here's how it goes .

The vulnerability constraint decoding method proposed by Andre Storhaug effectively avoids the security vulnerability in the auto-completion of smart contract code by embedding the vulnerability tagLLM in it\cite{70}. Bo Gao's research shows that ChatGPT is comparable to traditional tools in detecting machine unauditable vulnerabilities\cite{71}, but is easier to use and provides an evaluation data set. Wei Ma's iAudit framework combines fine-tuning and LLMs proxy, mimicking the human expert audit process to provide audit explanations that achieve high F1 scores and accuracy\cite{72}. Zhiyuan Wei's FTSmartAudit framework improves the efficiency and accuracy of smart contract audits through domain knowledge distillation and adaptive learning strategies\cite{73}. In addition, the AuditGPT tool and context-driven prompting technology further LLMs enhance the application in smart contract auditing\cite{74}. The publication of the SC-Bench dataset provides a rich resource for automated audit research\cite{75}. CoT prompts guideLLMs in-depth reasoning, improve audit accuracy, and jointly promote the development of smart contract audit technology\cite{76}.

  The development of these researches and tools not only provides new perspectives and solutions for the security audit of smart contracts, but also makes an important contribution to the safety, compliance, and trustworthiness of the entire blockchain ecosystem, ensuring that smart contracts play an even more crucial role in the digital world of the future.

\subsection{Vulnerablilities detection}
Smart contract audits are a core part of ensuring the security and reliability of blockchain technology. It not only effectively identifies and remediates potential security vulnerabilities and safeguards the integrity of user assets and the blockchain ecosystem, but also ensures that smart contracts comply with relevant legal and regulatory requirements, enhancing trust for users and participants. In addition, smart contract audits provide strong support for the development of blockchain technology by optimizing contract performance, promoting technological innovation, and academic research. It is also an important part of risk management, helping to reduce project risks and protect the interests of investors 

Recent research has proposed a variety of innovative approaches in the field of smart contract auditing, which aim to improve the accuracy and efficiency of auditing by combining the natural language processing and code understanding capabilities of large language models . The researchers have developed specialized frameworks and tools, such as FTSmartAudit \cite{77}, which fine-tunes LLMs to specialize smart contract audit tasks and utilizes domain-specific knowledge distillation techniques to generate high-quality datasets, while employing adaptive learning strategies to maintain the accuracy and robustness of the model. The SMARTSYS system optimizes hybrid dynamic analysis techniques through an interactive \cite{78}, self-decision-making approach, using predictive models to guide the use of fuzz testing to reveal deep vulnerabilities. In addition, some studies have proposed a vulnerability constraint decoding method to reduce LLMs the vulnerable code generated, and avoid the generation of vulnerabilities by embedding vulnerability tags and rejecting these tags at the decoding stage. The LLMSmartSec audit tool combines LLMs and annotates control flow diagrams \cite{79}, fine-tunes the model to understand the Solidity language, and analyzes smart contracts from different perspectives to identify and remediate vulnerabilities. The iAudit framework mimics the intuitive judgment process of human auditors \cite{80}, using fine-tuned Detector and Reasoner models to identify and analyze the cause of vulnerabilities. The SmartAudit framework implements a virtual smart contract audit organization through a multi-agent dialogue approach \cite{81}, and uses specialized agent roles to perform in-depth security analysis. LLM4FUZZ method is using LLMs to guide fuzz testing \cite{82}, optimize and automate smart contract security analysis, and effectively explore the code area by extracting the hierarchical representation of smart contracts and embedding them into the fuzz testing scheduler, so as to improve test coverage and vulnerability detection efficiency. These innovative approaches demonstrate LLMs the potential in smart contract audits, providing new ways to improve the security of smart contracts.

By combining the latest AI technologies with the specific needs of smart contract auditing, it provides new perspectives and tools to improve the security of smart contracts. As the LLMs technology continues to evolve and improve, it is expected that these methods will play a greater role in the field of smart contract security in the future. The results of these studies have brought substantial progress in the field of security auditing of smart contracts and provided a solid foundation for future research and development. 
\subsection{Language interaction and decision support in decentralized autonomous organizations (DAO)}
In the governance field of decentralized autonomous organizations (DAOs), the integration of large language models  has brought significant progress in automating the processing of complex data classification tasks, improving the efficiency and accuracy of DAO proposal classification, while highlighting the necessity of enhancing governance transparency and consistency. DAOs utilize blockchain technology to achieve automated management of the organization and collective decision-making among members through language interaction and decision support systems, enabling members to participate in the key decisions of the organization. 

Integrating the large language model LLMs into the decentralized autonomous organization (DAO) field has made significant progress in the automation of complex data classification tasks, especially in the context of DAO governance. A recent study by Ziegler and others has shown that LLMs automatically classifying DAO proposals is effective \cite{83}, a task that traditionally requires human expertise and is costly. Through iterative methods, researchers have developed a classification system with an accuracy of 95\%, highlighting the potential to fundamentally change data tagging tasks that rely on text context LLMs. At the same time, the governance process within DAOs is under scrutiny due to its vulnerability to attacks, leading to significant economic losses. Ma et al. conducted a comprehensive analysis of the DAO governance process \cite{84}, identifying issues in three core components: governance contracts, documents, and proposals. The authors constructed a state-of-the-art dataset containing 16,427 DAOs and developed automated methods to detect discrepancies, revealing significant gaps in transparency and consistency in the reviewed proposals. These studies collectively contribute to the theoretical and practical understanding of LLMs decision support systems and DAO governance. They provide tested DAO proposal classifications and insights into the design and engineering of decision support systems LLMs. The research results indicate that widespread adoption LLMs could lead to more effective and accurate classification of DAO proposals, while also emphasizing the necessity of improving the transparency and consistency of the governance process. Future research in this field may focus on evaluating the effectiveness of each proposal category, identifying bottlenecks in DAO governance, and further exploring their LLMs impact on individual and organizational decision-making processes.

\section{The Current Development Status Of The Combination Of Blockchain And Large Language Models}
 In today's era of rapid technological advancement, the integration of blockchain and large language models has become a high-profile focus in the field of technological innovation. Blockchain brings a new model to data storage and management with its characteristics of decentralization, non-tamperability, safety and reliability; large language model, with its powerful language understanding and generation capabilities, has demonstrated outstanding performance in natural language processing. Huge potential. However, the combination of the two is still in the stage of continuous exploration and development. Its application in smart contract development, data analysis, user interaction, security and other aspects has not only achieved remarkable results, but also faces many challenges. and urgent problems to be solved. Next, let us delve into the development status of the combination of blockchain and large language models.

\subsection{In Terms Of Smart Contract Development And Management}

\subsubsection{Code Generation And Auxiliary Development}

Although the discussion of the application of generative artificial intelligence in the construction industry focuses on specific areas, its ideas can bring inspiration to the code generation and auxiliary development of blockchain smart contracts\cite{1}. For example, its experience in handling complex tasks and providing innovative solutions helps achieve faster code generation, more accurate function implementation, and more innovative contract designs in the blockchain field. At the same time, using large language models and context learning to automatically generate smart contract annotations \cite{2} can improve the judgment and maintainability of the code. The large language model can automatically generate accurate and detailed comments based on the context of the contract code, improving development efficiency.

\subsubsection{Code Vulnerability Detection And Repair Suggestions}

The large language model plays a new role in intelligent vulnerability detection\cite{3}. It can break through the limitations of traditional vulnerability detection methods and provide more comprehensive and accurate detection services and vulnerability repair suggestions by learning a large number of rich codes and vulnerability cases. . In the field of smart contract code generation, there is an important method \cite{4} that can effectively avoid loopholes. This method is based on specific decoding technology, and its principle is to accurately control the code generation process by setting constraints related to vulnerabilities. In practical applications, many cases have emerged showing that this method can significantly improve the security of smart contracts, providing a strong guarantee for the steady development of smart contracts in blockchain and other related fields.

\subsection{In Terms Of Blockchain Data Analysis And Interpretation}
\subsubsection{Transaction Data Analysis}

Domain-specific large language models are of great significance in cryptocurrency operations\cite{5} and can handle the characteristics and challenges of its transaction data, providing insights such as price predictions, market trend analysis, and risk assessment.
The integration of artificial intelligence, Internet of Things, big data and blockchain is a key direction\cite{6}. Artificial intelligence and big data technology can process blockchain transaction data to mine value, while IoT devices expand paths for data collection and interaction, promoting collaborative technological innovation and application expansion.

\subsubsection{Smart Contract Audit}
A semantic-aware security audit method S-gram\cite{7} for Ethereum smart contracts was proposed. This method highlights the importance of semantic awareness, and uses case analysis to analyze its technical principles and implementation methods in detail, providing smart contracts with Security audits open up new avenues.

\subsection{In Terms Of User Interaction And Experience Of Blockchain Applications}
\subsubsection{Intelligent Customer Service And User Support}
Today, privacy protection large language models have attracted much attention. Taking ChatGPT as an example, privacy protection \cite{8} in intelligent customer service applications is crucial and faces challenges. Effective ways to protect privacy during service need to be explored.
At the same time, the development of Q\&A chatbot \cite{9} in the blockchain field is also advancing. By sorting out user needs and problem characteristics, the development process and technical implementation path are clarified, and its performance and user satisfaction are further evaluated to promote optimization and improvement.

\subsubsection{Personalized Services And Recommendations}

A streaming data interpretable recommendation method that integrates blockchain and large language models appears. It builds a recommendation mechanism with the help of blockchain characteristics and large language model capabilities. Its functions include data processing and explanation of recommendation reasons. The principle is that the two work together to ensure data and mining semantics. There are relevant cases to prove that it can improve recommendation accuracy and user trust.
In the field of financial technology, the combination of blockchain and large language models revolutionizes user experience/user interface design\cite{11}. Blockchain protects privacy, and the large language model understands needs. Its advantages are responding to personalized demands and creating intelligent interfaces. There are already applications to improve user satisfaction and competitiveness.

\subsection{In Terms Of Blockchain Security And Risk Management}
\subsubsection{Malicious Behavior Detection} 

In related research fields, the blockchain large language model provides a new response to malicious behavior detection \cite{12}, clearly introduces the specific detection methods used, and also objectively discusses the current limitations. and future development trends.
At the same time, a systematic literature review of large language models for network security\cite{13} also revealed many key points. It plays an important role in the field of network security, especially in the field of malicious behavior detection, effectively demonstrated its potential, and comprehensively discussed the subsequent research directions and many challenges faced.

\subsubsection{Smart Contract Security Enhancement}

A method to inspect smart contracts with the help of structural code embedding\cite{14} is proposed. The functions and principles of this method have been explained, and its effectiveness has been confirmed by cases, providing a new way to ensure the quality of smart contracts.
At the same time, in the research on the application of large language models in prediction and anomaly detection, a systematic literature review \cite{15} shows that it plays a significant role in the field of smart contract security enhancement and fully demonstrates its application potential. Related research directions and challenges It has also been discussed in depth, helping to further expand the boundaries of smart contract security technology.

\section{TECHNICAL ADVANTAGES}
In the current era of rapid technological development, blockchain and large language models each have their own advantages. Blockchain has the characteristics of decentralization, non-tamperability and traceability, ensuring data security and trust. Large language models are known for their powerful language processing capabilities and can understand natural language and generate high-quality text. When the two are combined, the blockchain provides secure data storage and verification for the large language model, and the large language model brings a more intelligent interaction method to the blockchain, jointly bringing efficient, safe and intelligent solutions to various fields.

\subsection{Large Language Models}
A large language model is a language model constructed from a deep neural network containing tens of billions of parameters. It is usually trained on a large amount of unlabeled text using self-supervised learning methods, such as Internet web page text, news articles, etc. Its technical architecture is mostly based on Transformer, with its attention mechanism, can learn vocabulary, syntax, semantic rules and contextual relationships, thereby having the ability to generate natural language text.

\subsubsection{Excellent Language Understanding And Production Abilities}
Large language models perform unsupervised learning through large-scale corpora, thereby in-depth mastering the grammar, semantics and pragmatic rules of natural language. Large language models represented by GPT-3 were fully demonstrated in OpenAI's research "Language Models are Few-Shot Learners". \cite{16} Through unsupervised learning of large-scale corpora, this model can accurately analyze complex language structures and semantic relationships like a scholar proficient in language. For example, in the scenario of few-shot learning, GPT-3 can quickly understand the task with only a few examples and generate high-quality, logically rigorous natural language text. Its language understanding and generation capabilities are amazing.

\subsubsection{Wide Range Of Application Scenarios}
Large language models have shown strong application potential in many fields. In terms of intelligent customer service, it is like a tireless and responsive "customer service elf" who can quickly and accurately respond to customers' various questions, greatly improving customer satisfaction. For example, many e-commerce platforms use intelligent customer service systems driven by large language models \cite{17} to efficiently handle massive user inquiries, which not only improves service efficiency but also reduces labor costs.

In the field of automatic text generation, large language models can create high-quality text content based on specific topics and requirements. In news reporting, it can quickly generate the first draft of news articles, reducing the work burden for reporters and increasing the speed of news output; in terms of novel creation, \cite{18} it can provide writers with rich creativity and inspiration, and even generate wonderful novel chapters to expand literature. The boundaries of creation.

In the field of machine translation, \cite{19} large language models can achieve more accurate and natural language conversion, effectively improving the efficiency of cross-language communication. Whether it is international business negotiations or academic research cooperation, it plays an important role.

In addition, in the field of education, \cite{20} large language models can serve as intelligent tutoring assistants to answer students' doubts and give learning suggestions. In the field of software development, it can assist programmers in generating code snippets and speed up the development process. In the field of marketing, accurate marketing copy can be generated based on user needs and market trends to improve marketing effectiveness. In the legal field, it can help lawyers quickly analyze a large number of legal documents, extract key information, and improve work efficiency.

\subsubsection{The Ability To Process Massive Data} 
Large language models learn and extract knowledge from large-scale data and can process massive amounts of text information. Using deep learning algorithms, models can automatically discover patterns and regularities in data and apply them to new tasks and problems. In the era of big data, in the face of massive text data, big language models can efficiently analyze and process, providing users with valuable information and profound insights. As proposed in "Attention Is All You Need", \cite{21} the Transformer architecture allows the model to process natural language more efficiently, better capture long-distance dependencies in text, and provides a key to the ability of large language models to process massive data. Technical support.

\subsection{Blockchain}
 
The concept of blockchain was first proposed by Satoshi Nakamoto in his 2008 paper "Bitcoin: A Peer-to-Peer Electronic Cash System."\cite{22} Blockchain is a distributed ledger technology that changes the way data is stored and managed through features such as decentralization, non-tamperability, security, reliability and traceability.

\subsubsection{Decentralized Features}

Blockchain relies on distributed ledger technology to completely get rid of the shackles of centralized control institutions, like a digital wilderness without absolute authority. Satoshi Nakamoto pointed out that "Bitcoin is a purely peer-to-peer version of electronic cash would allow online payments to be sent directly from one party to another without going through a financial institution."\cite{22} In the Bitcoin network, many nodes are like fearless nodes. The pioneers jointly guard the ledger, significantly reducing the possibility of single points of failure. In this distributed structure, every participant participates in decision-making and verifies transactions on an equal footing, effectively ensuring fairness and transparency.

\subsubsection{Immutability}

Blockchain creates an unchangeable and solid barrier for data through advanced encryption technology and consensus mechanism, which is like creating an unbreakable "digital fortress" for data. "An Overview on Smart Contracts: Challenges, Advances and Platforms"\cite{23} clearly states that "Smart contracts on blockchain benefit from the immutability and traceability of blockchain, ensuring the execution of contracts in a trusted environment." The hash function creates a unique identifier for the data , once the data is recorded, any modification will cause a change in the hash value, making it easily detectable. Taking financial transaction records as an example, each transaction is encrypted and closely linked to the previous transaction, forming an indestructible chain that cannot be tampered with.

\subsubsection{Safety And Reliability}
Blockchain uses complex encryption algorithms to encrypt data. Only users with the corresponding key can open this mysterious "digital door" and access and interpret the data. "Blockchain for Internet of Things: A Survey"\cite{24} emphasizes that "Blockchain provides decentralization, immutability, security, and traceability, which can address many challenges in the Internet of Things." In supply chain finance, blockchain is like a loyal A reliable "digital guardian" can effectively protect corporate business secrets and transaction information, resist data leaks and malicious attacks, and effectively protect user privacy and data security.

\subsubsection{Traceability}
Every transaction and data in the blockchain is recorded in chronological order and related to each other through a chain structure, just like weaving a fine "digital latitude and longitude network." "Blockchain-Based Secure Traceable Scheme for Food Supply Chain"\cite{25} takes the food supply chain as an example, "The blockchain-based traceability system can provide consumers with transparent and reliable information about the food supply chain." By scanning the product QR code, consumers can It's like having a "digital microscope" that can clearly understand the product's raw material sources, production processes, transportation routes and other information to ensure product quality and safety.

\subsection{Combining Blockchain And Large Language Models}
The combination of large language models and blockchain has many potentials. On the one hand, the non-tamperable and traceable features of blockchain can be used to verify the source of training data and generated results of large language models, improving credibility; on the other hand, large language models It can assist in generating smart contract code and logic, and optimize smart contract interaction. The combination of the two brings innovative applications and advantages to multiple fields.
\subsubsection{Security Advantages}

With its powerful language understanding and analysis capabilities, the large language model can accurately identify potential vulnerabilities and risks in smart contracts and provide effective suggestions for vulnerability repair, thereby improving the security of blockchain transactions \cite{26}. For example, during the writing process of smart contracts, large language models can conduct detailed analysis of the code to detect possible logic errors and security risks.

The large language model can also quickly and accurately identify abnormal transaction behaviors \cite{26}, providing real-time security protection for the blockchain network. By learning and analyzing large amounts of transaction data, big language models can build models of normal trading patterns. Once a transaction occurs that is inconsistent with the model, an alert can be issued in a timely manner.

\subsubsection{Governance Advantages}
Large language models can achieve decentralization and collaborative governance in a blockchain environment \cite{27}. The decentralized nature of the blockchain provides a new application platform for large language models, allowing different participants to collaborate, use large language models to analyze various types of data on the blockchain, and mine valuable information for the blockchain. Governance provides the basis for decision-making. For example, the large language model can conduct in-depth analysis of the discussion content of the blockchain community, extract key issues and suggestions, promote the participation and collaboration of community members, and improve the efficiency and transparency of governance. At the same time, the model enhances data security, improves the fairness and transparency of the model, and provides new ideas for the integrated development of artificial intelligence and blockchain.

\subsubsection{Unique Value And Irreplaceability}
 
The decentralized and non-tamperable characteristics of blockchain provide a reliable data storage and verification mechanism for large language models to ensure that the data is authentic and trustworthy. The document "Blockchain convergence: Analysis of issues affecting IoT, AI and blockchain" \cite{28} comprehensively discusses the convergence trends and future development directions of blockchain and multiple technologies including the field of artificial intelligence to which large language models belong. It is essential for understanding blockchain The uniqueness combined with large language models has important reference value. It provides a detailed analysis of how the characteristics of blockchain provide a unique data foundation and verification mechanism for large language models.

At the same time, the large language model brings innovation and expansion to blockchain application scenarios, such as smart contract optimization and blockchain social platform content management. It can be seen in the literature that the powerful language processing capabilities of large language models can provide more possibilities and innovative ideas for various application scenarios of blockchain. The document "Intersection of AI and blockchain technology: Concerns and prospects"\cite{29} specifically focuses on the intersection of artificial intelligence and blockchain technology, expounding the opportunities and challenges brought by the combination of the two. Among them, large language models are an advanced application of artificial intelligence , its synergy with blockchain is also reflected.

According to the document "Revolutionizing cyber threat detection with large language models: A privacy-preserving bert-based lightweight model for iot/iiot devices", \cite{30} although it mainly focuses on IoT security, the concept can be extended to the blockchain field. Large language models can improve the ability to identify complex network attacks and provide reference for blockchain security protection. At the same time, this model provides a new solution to the problems of limited resources and privacy protection in blockchain while ensuring privacy protection.

The document "Blockchain smart contracts formalization: Approaches and challenges to address vulnerabilities" \cite{31} focuses on the formalization method of blockchain smart contracts, providing a theoretical basis and practical guidance for improving the security of blockchain smart contracts. Large language models can be combined with these formal methods to better deal with vulnerability issues in smart contracts and improve the security of blockchain transactions.

\section{DEVELOPMENT CONSTRAINTS}
Both large language models (LLMs) and blockchain technologies face significant constraints in their development. This section explores the key challenges for both fields.
\subsection{Large Language Models}
Large language models play a crucial role in modern artificial intelligence but encounter several challenges in real-world applications:
\subsubsection{Computational Resources}
LLMs typically require vast computational resources for both training and inference. This demand for resources limits the widespread adoption of these models, especially in environments with constrained computing power.

\subsubsection{Explainability and Transparency}
As LLMs grow in complexity, Explainability and transparency become essential. Understanding the decision-making process of these models is critical to ensure that their outputs are reasonable and trustworthy.

\subsubsection{Data Quality and Bias}
The quality of training data directly affects the performance of LLMs. Biases or noise in the data can lead to inaccurate or biased outputs, which must be carefully managed.

\subsubsection{Ethics and Privacy}
LLMs rely on vast amounts of user data. Ensuring that this data is handled ethically and that privacy concerns are addressed is a key constraint in the development of these models.

\subsubsection{Training Data Security}
The security of training data is vital for the robustness of LLMs. Data attacks, such as poisoning attacks and adversarial samples, can significantly impact the model's performance.

\paragraph{Poisoning Attacks}
In poisoning attacks, malicious actors inject harmful samples into the training data, causing the model to produce incorrect results during inference.

\paragraph{Adversarial Sample Attacks}
Adversarial sample attacks involve small modifications to input data that lead to significant changes in the model's predictions. This type of attack poses a substantial threat to the security of LLMs.

\subsection{Blockchain}
Blockchain technology also faces numerous challenges in its development, which are critical for ensuring its continued viability and growth:

\subsubsection{Security}
Security is one of the most fundamental features of blockchain systems. Any security breach can lead to the collapse of the system or a loss of trust among users. Blockchain must ensure that transactions and records are tamper-proof, and various consensus algorithms (like Proof of Work and Proof of Stake) are in place to prevent attacks such as double-spending or malicious block validation.

\subsubsection{Privacy}
In a blockchain network, privacy is a significant concern. The transparency of public ledgers makes users' transaction records visible, posing a challenge to maintaining user privacy while ensuring transparency. Several privacy-preserving techniques, such as zk-SNARKs (zero-knowledge succinct non-interactive arguments of knowledge) and ring signatures, are being explored to enhance blockchain privacy without compromising on security.

\subsubsection{Latency}
Blockchain systems often struggle with transaction confirmation times, which affects their usability in real-time applications such as payments or high-frequency trading. Blockchain transactions typically take longer to be confirmed compared to traditional centralized systems. Improving the system's latency is a major challenge, and researchers are working on solutions like off-chain transactions, Layer 2 solutions (e.g., Lightning Network), and faster consensus algorithms to speed up the process.

\subsubsection{Scalability}
Scalability in blockchain refers to improving the system's capacity to handle a growing number of transactions while maintaining decentralization and security. As blockchain networks grow, they face challenges related to the volume of transactions and the storage requirements of maintaining the entire transaction history.

\paragraph{Transaction Throughput}
Blockchain networks are often limited by the number of transactions they can process per second (TPS). Current networks, such as Bitcoin and Ethereum, can handle only a few transactions per second, whereas traditional payment systems like Visa can handle thousands. To improve throughput, various solutions like sharding, increased block sizes, and more efficient consensus mechanisms are being explored.

\paragraph{Sharding Techniques}
Sharding is a technique aimed at improving scalability by distributing the transaction load across multiple nodes. By dividing the blockchain into smaller pieces or "shards," each containing a subset of the network's data, sharding can reduce congestion and improve processing times. However, implementing sharding presents complex challenges, such as ensuring data consistency and security across all shards.

\paragraph{Chain Storage}
As blockchains continue to grow, the length of the chain increases. Efficiently storing and accessing historical data is an ongoing challenge for blockchain systems. Storing the entire transaction history on every node can lead to storage bloat, making it harder for new participants to join the network. Solutions such as pruning (removing old transactions from nodes) and the use of off-chain data storage mechanisms are being researched to address this issue.

\section{PROSPECTS}
  The combination of blockchain technology and large-scale language models indicates great potential for future development. Recent developments in federated learning and cloud robotics~\cite{liu2019lifelong,liu2019federated,liu2021peer} have demonstrated the potential for distributed learning systems, with successful applications in healthcare~\cite{liu2020experiments} and real-time contribution measurement~\cite{yan2021fedcm}. Advances in elastic and secure robotic systems~\cite{liu2022elasticros,zhang2022authros,liu2023roboec2} offer promising directions for enhancing distributed AI capabilities. The applications in smart cities~\cite{zheng2022applications}, traffic prediction~\cite{liu2017singular}, and autonomous driving~\cite{liu2024edgeloc} showcase the practical value of these technologies. Recent explorations in network slicing for large language models~\cite{liu2024llm} further highlight the potential for specialized infrastructure to support AI systems. Blockchain, as a distributed ledger technology, enables secure and trusted transactions without the involvement of a third party. Its characteristics of decentralization, immutability and transparency, combined with the powerful language processing capabilities of large language models, are expected to lead to revolutionary changes in various industries. This convergence technology is particularly suitable for those areas that usually require a lot of third-party authentication, transaction costs and trust issues, such as government, finance, banking, etc. \cite{alam2021blockchain}. In the next decade, the Industrial Internet of Things will be critical to the development of the social Internet of Things, aiming to create a sustainable and secure green edge infrastructure for iot applications. This includes social IIoT systems, affordable and clean energy,green communication and computation, fog/edge infrastructure, and cybersecurity. In order to improve network efficiency and ensure network security, these areas will need to rely on advanced large-scale language models and blockchain technology \cite{hazra2022blockchain}. In the future, areas such as government, health, finance, economics and energy are expected to adopt this convergence technology. This fusion technology will have broad application prospects in the future.

\section{CONCLUSION}
  This review provides a summary of the current impact between blockchain and LLMs. In order to better reflect the relationship between blockchain and LLMs, we summarize the advantages and disadvantages of blockchain and LLMs, as well as examples of how each has been used in the other's domain. We hope that this review will give readers a better understanding of the current situation of blockchain and LLMs, and contribute to the future exploration of this area.

\bibliographystyle{ieeetr}
\bibliography{references.bib}
\end{document}